\documentclass[aps,prb,superscriptaddress,twocolumn]{revtex4-1}
\usepackage{macros_BW}

\begin{document}

\title{High-Pressure Elasticity Tensor from Thermal Diffuse Scattering}

\author{Bj\"orn Wehinger}
\email[]{bjorn.wehinger@unige.ch}
\affiliation{Department of Quantum Matter Physics, University of Geneva, 24, Quai Ernest Ansermet, CH-1211 Gen\`eve, Switzerland}

\author{Alessandro Mirone}
\email[]{mirone@esrf.fr}
\affiliation{European Synchrotron Radiation Facility, 71, Avenue des Martyrs, F-38000 Grenoble, France}

\author{Michael Krisch}
\affiliation{European Synchrotron Radiation Facility, 71, Avenue des Martyrs, F-38000 Grenoble, France}

\begin{abstract}
We report on the determination of high-pressure elasticity from thermal diffuse x-ray scattering of magnesium oxide at pressures up to 28 GPa. We find that the full elasticity tensor as a function of pressure can be extracted from x-ray scattering of single crystals. This enables the simultaneous measurement of elasticity and crystal structure from a single experiment and thus marks a decisive step towards establishing reliable calibration for extreme pressures. Elastic constants can be extracted with this method from small crystals of arbitrary symmetry, shape and optical properties and will allow for significant progress in our understanding of the elastic behaviour of planets interior at geophysically relevant conditions, structural phase transitions and fundamental interactions of phonons with other quasi-particles.
\end{abstract}

\maketitle

\section{Introduction}
Accurate pressure calibration is a central topic in high-pressure science and a challenging problem for static pressure experiments at several hundred GPa. State-of-the-art experiments above 100\,GPa static pressure can be conducted on single crystals and provide precise results on the crystal structure and equation of state~\cite{mcmahon_structure_2007,dewaele_equations_2004}. The combined use of small crystals and helium loaded diamond anvil cells assures excellent hydrostaticity and enables high-precision measurements using x-ray diffraction. The limiting factor for even higher accuracy in such experiments often is the uncertainty on absolute pressure. While shock-wave experiments allow for a direct measurement of pressure and provide equation of states with an accuracy of 1-3\,\% at high temperatures~\cite{yokoo_ultrahigh-pressure_2009,smith_ramp_2014}, static pressure experiments using diamond anvil cells require the use of a secondary pressure scale that is calibrated against shock-wave experiments~\cite{sokolova_self-consistent_2013}. Highly accurate calibration is limited by the fact, that the temperatures used in static pressure experiments are often very different from shock-wave experiments which in turn requires the use of additional approximations for the calibration. Furthermore, using a secondary pressure scale requires the pressure calibrant to be at exactly the same conditions as the sample itself. This becomes experimentally very challenging and critical for pressures above 100\,GPa where the volume on which excellent hydrostaticity is guarantied becomes very small. Establishing a primary pressure scale is thus desirable. Such can be obtained by the combined use of structural and elastic properties and direct calibration up to 55 GPa was obtained from magnesium oxide by combining structural data from x-ray diffraction with the elasticity tensor derived from Brillouin light scattering~\cite{zha_elasticity_2000}. The obtained accuracy is rather good (about 1\,\%) and helped to improve the calibration of ruby fluorescence. The drawbacks are the difficulties of extending high-pressure Brillouin light scattering to opaque materials and the complexity of the combined experimental setup. A suitable alternative for the high-pressure measurement of elastic properties is inelastic x-ray scattering that can be applied to high temperatures as well~\cite{antonangeli_elasticity_2008}. However, such experiments require very sophisticated spectrometers and are very time consuming for systems of low symmetry, where different orientations of the crystal need to be applied. Recent progress on the quantitative analysis of thermal diffuse scattering (TDS) shows that the full elasticity tensor can be obtained from x-ray scattering on single crystals~\cite{wehinger_full_2017}. This allows for the simultaneous measurement of the crystal structure and elasticity in a single experiment. 

The information on the full elasticity tensor at geophysical relevant pressures furthermore contains key information for connecting sound velocities of minerals to seismic wave propagation which in-turn allows for conclusive statements on  conditions and composition of the interior of planets~\cite{antonangeli_composition_2010,wehinger_dynamical_2016}. In condensed matter physics, pressure-induced changes in the elasticity tensor capture subtle details of the interplay of phonons with other quasiparticles. Electron-phonon and magnon-phonon coupling is visible in thermal diffuse x-ray scattering~\cite{bosak_TDS_2009,tacon_inelastic_2014,toth_electromagnon_2016} and important for electronic topological phase transitions, high-temperature superconductivity and quantum magnetism. Precise measurements of the elasticity tensor in the vicinity of structural phase transitions allow for a detailed understanding of the underlying mechanism. This is in particular true for high-pressure phase transitions, where the treatment of finite strain leads to a consistent Landau theory~\cite{troster_fully_2014}.

In this study, we show that the full elasticity tensor can be obtained from x-ray scattering experiments at high pressures, thus enabling the establishment of a primary pressure scale from a single measurement for static pressures. Our model-free analysis of TDS allows for experiments on single crystals of arbitrary symmetry and shape regardless of optical properties. We benchmark our approach on magnesium oxide for static pressures up to 28\,GPa using a diamond anvil cell.

\section{Results}

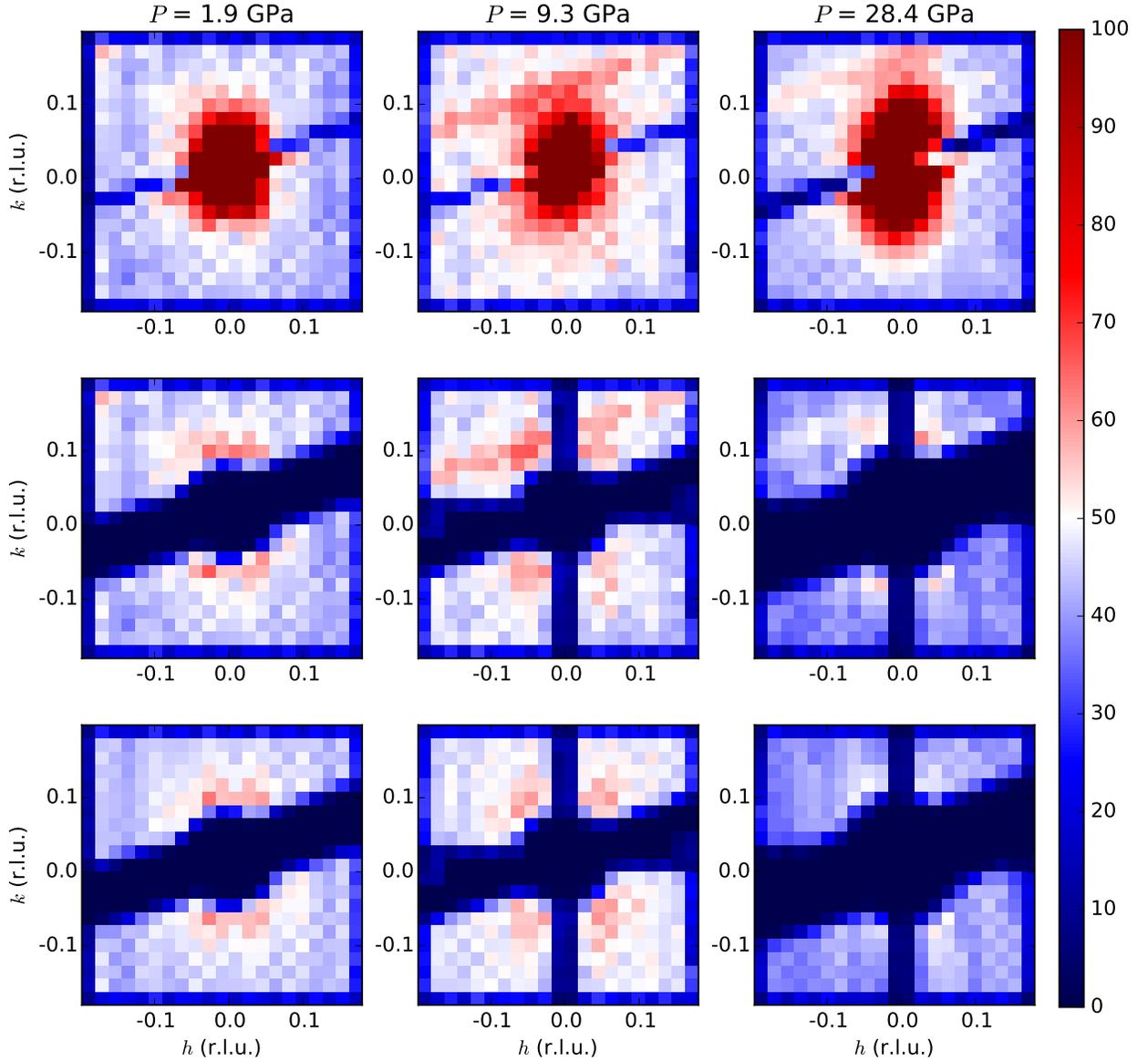
\figuremacroW{MgO-high-pressure-diffuse}{Diffuse scattering of magnesium oxide at high pressure and room temperature. Top panels: Reconstructed experimental diffuse scattering in the vicinity of the $[4\,0\,0]$ reflection at indicated external pressure and $l$ = 0.126\,r.l.u. Plotted intensities are obtained from the irregular experimental grid by inverted interpolation. The Bragg reflection saturates the detector on one diffraction pattern for $P$ = 1.9 and 9.3\,GPa and two for 28.4\,GPa. These patterns were excluded (dark blue area). Middle panels: Analyzed part of experimental diffuse scattering for the same pressures and same $l$. Areas with zero intensity (dark blue) are affected by elastic diffuse scattering and masked. Bottom panels: Calculated TDS from the obtained elastic moduli.}{1.0}

TDS from magnesium oxide was measured at the high-pressure beamline ID27 of the European Synchrotron Radiation Facility. High-quality samples were prepared by mechanical cutting and gentle etching. A single crystal of size 50$\times$50$\times$20\,$\mu$m$^3$ was placed in a membrane driven diamond anvil cell with 70$^\circ$ opening. Natural diamonds of 1.6\,mm thickness and culets of 450\,$\mu$m in diameter were used with rhenium as gasket. Helium served as pressure medium to ensure optimal hydrostaticity and minimal contribution to the scattering. Fluorescence from two ruby crystals was used for independent pressure measurement. Monochromatized x\,rays of wavelength $\lambda$ = 0.62\,\AA\xspace and a spot size of 14$\times$20\,$\mu$m$^2$ were used and scattering intensities were collected in transmission geometry with a noise-free single photon counting hybrid detector equipped with 300\,$\mu$m thick silicon pixels of size 172$\times$172\,$\mu$m$^2$ (PILATUS-300K from Dectris, Switzerland) placed 130.291\,mm behind the sample. The cell was rotated orthogonal to the incoming x\,rays with angular steps of 0.1$^\circ$ exploiting the full angular range of the cell opening. In order to insure a monotonous absorption of scattering from the diamonds we first performed a careful absorption measurement to determine the angular ranges for which neither incoming nor scattered x\,rays satisfied the Bragg condition of the diamonds. The diffraction geometry, orientation matrix and cell volume were refined by a Fourier analysis of all collected magnesium oxide Bragg reflections. 
Experimental scattering intensities in proximity to selected Bragg reflection measured at room temperatures and pressures up to 28\,GPa are shown in the top panels of Fig.\,\ref{fig:MgO-high-pressure-diffuse}. Scattering intensities decay smoothly with increasing $q$. The anisotropic shape of the diffuse scattering reflects the anisotropy in the elastic tensor: Scattering intensities decrease faster in $q$ for longitudinal (horizontal) compared to transverse (vertical) momentum transfers. The shape of the diffuse scattering remains similar with increasing pressure despite a broadening of the Bragg reflections due to increasing internal stresses. 
For fitting the four rank elasticity tensor $c_{ijlm}$, we have selected two strong reflections with minimal contribution to elastic diffuse scattering. Data were analyzed using the recently established method for determination of the full elasticity tensor from TDS~\cite{wehinger_full_2017}. The reciprocal volume to be analyzed is limited to momentum transfers within an interval [$q_{min} , q_{max}$] according to the following conditions: First, the dispersion of the acoustic phonons dominating the TDS must fulfill the elastic approximation. We thus excluded the region with reduced wavevectors $q > 0.18$ reciprocal lattice units (r.l.u.) verified by \textit{ab initio} calculations at ambient conditions~\cite{wehinger_full_2017}. Second, to minimize the contribution of elastic scattering, we excluded regions very close to the Bragg reflections. For our high-pressure experiment we find the exclusion according $q < 0.04$\,r.l.u. appropriate for $P$ = 1.9\,GPa and 9.3\,GPa and $q < 0.05$\,r.l.u. for 28\,GPa. This choice is justified by iterative fits with decreasing $q_{min}$.
The observed scattering is the convolution of diffuse scattering from diamond and the sample. In our analysis we assume the contribution from the diamonds to be constant across the analyzed region of interests. This is justified by background measurements beside the crystal. In order to keep the influence of the convolution effect low, we furthermore exclude pixels with intensities of $> 1500$ counts and those on the same diffraction pattern within an exclusion radius of 40 pixels. The analyzed intensities are illustrated in the middle panels in Fig.\,\ref{fig:MgO-high-pressure-diffuse}. Diffuse scattering due to stacking faults in the sample contributes as well to the elastic part. Such contribution appears along high symmetry direction for the two higher pressure points and can be excluded by carefully masking the scattering along these directions in reciprocal space. Intensities along three main crystallographic directions were masked with slabs of thicknesses $d_h = d_k = d_l = 0.036$\,r.l.u. 

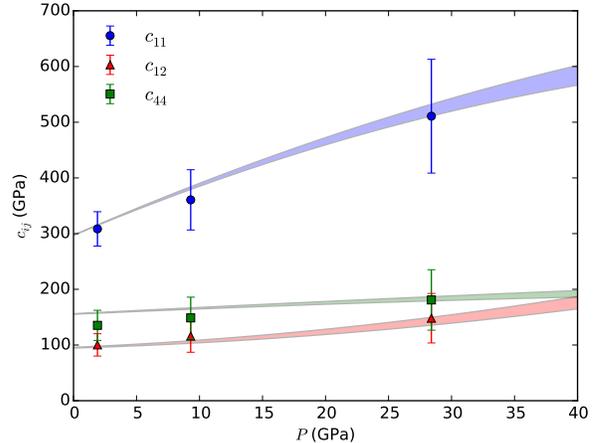
\figuremacro{MgO-high-pressure-elasticity}{Elastic moduli for magnesium oxide at room temperature as a function of external pressure. Symbols with error bars were obtained from fits to diffuse scattering intensities in the vicinity of reflections  $[2\,0\,0]$, and $[4\,0\,0]$. Filled area between solid lines are fits to Brilluoin scattering results from Ref.~\cite{zha_elasticity_2000}. Pressure was determined from ruby fluorescence.}{1.0}

For the fit we assume the validity of the adiabatic and harmonic approximations and compute scattering intensities for single phonon processes~\cite{xu_zkri_2005,bosak_jap_2015}. For the selected regions in reciprocal space the TDS is dominated by acoustic phonons which in turn follow the elastic wave equation
\begin{equation}
\label{eq:el_motion}
\rho \omega^2 u_{i} = c_{ijlm} k_j k_l u_m,
\end{equation}
with $\rho$ the mass density, $ \bs{k} = k \bs{n} $ the wave vector and $ \omega $ the frequency of the elastic waves in good approximation~\cite{fedorov_PP_1968}. Eq.\,(\ref{eq:el_motion}) is solved for the given crystal symmetry and $c_{ijlm}$ by solving the minimization problem 
\begin{equation}
\label{eq:monotemp}
c, {\bs{b}},{\bs{g}} =  \underset{c^\prime , {\bs{b}^\prime},{\bs{g}^\prime}}{\operatorname{argmin}} \left( \sum_{ {\bs{Q}}}  \left( I^{calc}_{{\bs{Q}}, T}( c^\prime, {\bs{b}^\prime} ,{\bs{g}^\prime}) -  I^{exp}_{{\bs{Q}}, T} \right)^2\right), 
\end{equation}  
where  $I^{exp}_{{\bs Q}, T} $ is the measured scattering intensity at temperature $T$ over a set of reciprocal space points  $ \bs{Q} \in \left\{ Q_{exp}  \right\}$ in the proximity to Bragg reflections, $I^{calc}_{{\bs{Q}}, T}$ is the calculated intensity.
The elastic tensor $c$, background $\bs{b}$ and renormalization factor $\bs{g}$ are the fit parameters. $\bs{b}$ and $\bs{g}$ are arrays that are kept constant in proximity to individual reflections, see Ref.~\cite{wehinger_full_2017} for details.
The diffuse scattering contribution from diamonds is treated constant in the vicinity of each Bragg reflection and taken into account by $\bs{b}$. Absorption, polarization and geometrical corrections for both, sample and diamonds are accounted for by $\bs{g}$. Scattering intensities on individual pixels were fitted simultaneously, with the exact diffraction geometry and lattice parameters constraint to the results of the Fourier analysis. TDS computed from the obtained elastic modui is illustrated in the bottom panels of Fig.\,\ref{fig:MgO-high-pressure-diffuse}.

By fitting high-pressure scattering intensities collected at a single temperatures we obtained the full elasticity tensor upon a single scaling factor. Absolute values of $c_{ijlm}$ are obtained by normalization to the bulk modulus for magnesium oxide~\cite{zha_elasticity_2000}. The pressure evolution of the elastic modulii is shown in Fig.\,\ref{fig:MgO-high-pressure-elasticity}, were we compare the results of our study to previous results obtained from fits to Brillouin scattering~\cite{zha_elasticity_2000}. The values obtained from fitting the diffuse scattering intensity distribution increase with pressure and follow the different slopes and curvatures as established previously within errors. 
The error bars on the elastic moduli increase with pressure. The reason is decreasing crystal quality which leads to stronger contribution of elastic diffuse scattering. 

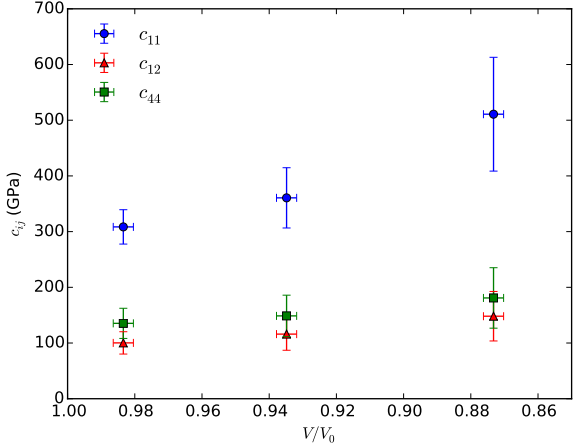
\figuremacro{MgO-elasticity-vol-dep}{Volume-dependence of the elastic moduli for magnesium oxide at room temperature. Symbols with error bars were obtained from fits to diffuse scattering intensities in the vicinity of the selected reflections. The cell volume was determined by Fourier analysis of the ensemble of collected Bragg reflections with an error of $\sigma_V/V = 6 \times 10^{-3}$.}{1.0}

In order to establish the volume-dependence of the elastic moduli we refine the unit cell volume at each pressure point by a Fourier analysis of the ensemble of collected magnesium oxide reflections. The result is shown in Fig.\,\ref{fig:MgO-elasticity-vol-dep}. 

\section{Discussion}
The rigorous treatment of diffuse scattering measured at high pressures using diamond anvil cells allows for the extraction of the full elasticity tensor. The fit is sensitive to the correct choice of the region in reciprocal space to be analyzed. The conditions for the momentum transfers to be analyzed and the regions in reciprocal space to be excluded were established by iterative fits and justified by \textit{ab initio} calculations at ambient pressure~\cite{wehinger_full_2017}. Diffuse scattering from the diamonds and sample environment was treated as constant contribution. This approximation is justified by the fact that the dominant Compton scattering from diamond and contribution from air scattering varies very little across the small regions of interest. Elastic diffuse scattering was excluded by careful masking of affected regions. Alternatively, small contributions of elastic diffuse scattering can be subtracted by exploiting the temperature dependence of TDS which differs significantly from the temperature dependence of elastic scattering of both the sample and environment. Analyzing the intensity differences measured at two or more slightly different temperatures can increase the precision of the measurements significantly and provides absolute values of the full elasticity tensor without scaling factor~\cite{wehinger_full_2017}. Such experiments can be performed at high pressures and we expect improvements to similar accuracy as obtained at ambient conditions. In measurements with increased statistics and momentum resolution it will be possible to probe small temperature differences which allow to study the evolution of the elasticity tensor over a wide range of temperatures. Determination of absolute values becomes relevant if the bulk modulus is unknown. 

The errorbars on the individual elastic moduli form the data of this pilot experiment are large. This is due to the limited data quality obtained in this experiment. Statistics and momentum resolution are rather poor, see upper panels in Fig.\,\ref{fig:MgO-high-pressure-diffuse}.
Both aspects can easily be improved in future experiments by longer counting times and an increased sample-detector distance. A significant improvement can be achieved by the use of a better suited detector. In fact, the use of a silicon based hybrid photon counting detector restricted our choice of the incident wavelength with drawbacks in both quantum efficiency and accessible reciprocal space. The small detection area furthermore required identical measurements at three vertically offset detector positions. A large area detector with small pixels size and cadmium telluride as detection material will overcome these issues due to better quantum efficiency at smaller wavelength, increased momentum resolution and extended reciprocal space to be analyzed. Higher statistics on the scattering intensities will furthermore allow to employ a de-convolution scheme for the treatment of additional contributions to diffuse scattering such as Compton scattering from diamonds and fluorescence. Such treatment will allow us to extend the conditions for the reciprocal space section to be analyzed to lower relative momentum transfers.

A primary pressure scale can be obtained from the volume-dependence of the elastic modulii (Fig.\,\ref{fig:MgO-elasticity-vol-dep}) using the thermodynamic relation 
\begin{equation}
\label{eq:td}
K_T = -V  \left( \frac{\partial P}{\partial V} \right)_T, 
\end{equation}
where $K_T$ is the isothermal bulk modulus which can be obtained from the elastic moduli via the Reuss-Voigt-
Hill relation~\cite{hill_pps_1952}. Assuming $K_T$ and $V$ to be measured as function of some empirical pressure parameter $\Pi$, one can integrate Eq.\,\ref{eq:td} along an isotherm. This yields the thermodynamic pressure $P$ with
\begin{equation}
\label{eq:P}
P(\Pi) = - \int^{\Pi}_{0} \frac{K_T}{V}\left( \frac{\partial V}{\partial \Pi} \right)_T d \Pi .
\end{equation}
$\Pi$ can thus be calibrated in terms of absolute pressure if $K_T$ and $V$ are measured over a wide range of pressure points. For estimating the accuracy of this approach using the proposed methodology we replace the integral in Eq.\,\ref{eq:P} by a sum 
\begin{equation}
\label{eq:Psum}
P = - \sum_{i=1}^{n} K_i \left(1 - \frac{V_{i-1}}{V_i} \right),
\end{equation}
where 
$K_i$ and $V_i$ are the $n$ individual measurements of $K_T$ and $V$ at discrete values of $\Pi$.
For closely spaced measurements ($V_{i-1}/V_i \approx 1$) and assuming $\delta K_i/K_i = \sigma_K/K$ and  $\delta V_i/V_i = \sigma_V/V$ the error on P can be approximated by
\begin{equation}
\label{eq:error}
\frac{\sigma P}{P} \approx \sqrt{ \left( \frac{ s_K}{K} \right)^2 +  \frac{1}{n}  \left(  \frac{\sigma_K}{K} \right)^2 + \frac{K_n^2 + K_0^2}{P^2}  \left( \frac{\sigma_V}{V} \right)^2 } ,
\end{equation}
where $s_K/K$ is the systematic error in our measurements due to the approximations in the determination of $K_T$ (Ref.~\cite{decker_proposed_1970}). Systematic error on $V$ cancel.
Assuming $s_K/K = 10^{-2}, \sigma_K/K = 10^{-2}$ and $\sigma_V/V = 10^{-3}$ at $n = 50$ equally spaced measurements we obtain $\sigma P/P = 10^{-2}$ at P = 100\,GPa for magnesium oxide, where  $s_K/K$ is the dominating contribution at high pressure. The assumption $\sigma_K/K = s_K/K = 10^{-2}$ is smaller than our result but a realistic estimate for future experiments realized at two or more temperatures per pressure point. The accuracy of the primary pressure scale might further be improved using very high quality crystals of lower symmetry, where the extraction of elastic moduli from TDS works remarkably well~\cite{wehinger_full_2017}. 
Our method can be extend not only to systems of low symmetry but also to opaque systems where measurements using Brillouin scattering get to their limits. The rather universal application of our method to small single crystals of arbitrary shape, symmetry and optical properties opens attractive possibilities for elasticity measurements at geophysical relevant pressures. Accurate measurements will help to improve the understanding of condition and composition of planets interior by comparison with seismic data. The quantitative treatment of TDS furthermore constrains the atomic displacement patterns~\cite{wahlberg_implications_2017} which is important for the crystallographic refinement of new high-pressure phases.

\section{Summary}
In summary, we have shown that the elastic moduli of a single crystal can be obtained by analyzing diffuse scattering in a high-pressure x-ray diffraction experiment. Elastic properties can be extracted together with the crystal structure at the same conditions. This is of outstanding relevance for high-pressure measurements in material science, geophysics and for the investigation of fundamental interactions of other quasi-particles with phonons in condensed matter physics. 
The simultaneous measurements allow to establish a primary pressure scale using a single experimental technique. This becomes relevant at high static pressures, where excellent hydrostatic conditions can only be assumed on a very small volume.

\section*{Acknowledgment}
We thank Volodymyr Svitlyk, Alexe\"i Bosak and Jeroen Jakobs for excellent support with the experiment and help in preparation of sample and pressure cells. Mohamed Mezouar and Alexe\"i Bosak are greatly acknowledged for fruitful discussions. The analysis software is available as open source code at \url{http://ftp.esrf.fr/scisoft/TDS2EL/}.

\bibliographystyle{apsrev4-1_BW}
\bibliography{high-pressure-elasticity}

\end{document}